# Revisiting the Role of Plasma Sheet Bubbles in Stormtime Energy Transport Using RCM-I

Sina Sadeghzadeh[1], Frank Toffoletto[1], Vassilis Angelopoulos[2], and Richard Wolf[1]

[1]Physics and Astronomy Department, Rice University, Houston, TX, USA
[2]Department of Earth, Planetary, and Space Sciences, UCLA, Los Angeles, CA, USA

## ABSTRACT

Plasma sheet bubbles, defined as entropy-depleted flux tubes, are widely regarded as an efficient mechanism for transporting plasma into the inner magnetosphere during geomagnetic storms. Equilibrium simulations using the Rice Convection Model (RCM-E) predict that bubbles can account for up to ~60% of storm-time ring current energy during intense storms. However, global simulations and observations suggest a more moderate net contribution. In this study, we quantify the contribution of plasma sheet bubbles to ring current buildup using a Lagrangian particle backtracking technique applied to three idealized storm simulations conducted with the inertialized Rice Convection Model (RCM-I). A stratified ensemble of ~100,000 test particles, weighted by local plasma pressure and entropy, was traced backward in time to determine whether their energy originated from bubble injections, non-bubble plasma sheet transport, or pre-existing trapped populations. Our results show that bubble contributions increase with storm intensity but saturate near ~40% of the total ring current energy inside $R < 6.6R_e$, even for strong storms (Dst ≈ −180 nT). The trapped population remains comparably important (~40%), while non-bubble transport contributes ~15%. This saturation is notably lower than the ~61% predicted by RCM-E and is attributed to inertial braking, which generates oscillatory flows and tailward return streams that remove approximately 40% of the inward bubble energy flux. When only newly transported plasma is considered, bubbles account for ~73% of the inward transport, consistent with global MHD and flux-based studies. These results reconcile equilibrium modeling, global simulations, and spacecraft observations by demonstrating that bubbles dominate inward transport but do not fully replace the resident ring current population due to inertial limitations.

## PLAIN LANGUAGE SUMMARY

During geomagnetic storms, plasma from the distant magnetosphere is transported inward and builds up the ring current, which weakens Earth's magnetic field. This transport often occurs in short-lived, fast-moving structures called plasma sheet "bubbles." Previous models suggested that bubbles could account for most of the storm-time ring current. Using a physics-based model that includes plasma inertia, we tracked the paths of many representative particles backward in time to determine where the ring current energy comes from. We find that bubbles become more important as storms intensify, but contribute only about 40% of the total ring current energy, even during strong storms. A similar fraction comes from plasma that was already trapped closer to Earth before the storm began. Our results show that while bubbles carry most of the newly injected energy, much of it is lost through return flows driven by inertial braking. This explains why spacecraft observations find smaller bubble contributions than earlier models predicted.

## KEY POINTS

- Plasma sheet bubble contributions to ring current energy increase with storm intensity but saturate near ~40% in inertial simulations.
- Inertial braking and return flows reduce the net energy retained from bubble injections compared to equilibrium model predictions.
- Bubbles dominate inward transport of new plasma (~70%), but pre-existing trapped populations remain a major component of storm-time ring current energy.

## INTRODUCTION

The buildup of the storm-time ring current is a defining feature of geomagnetic storms and represents a major channel through which solar wind energy is deposited into the inner magnetosphere. Traditionally, this buildup has been attributed to large-scale, quasi-steady convection driven by enhanced dawn-to-dusk electric fields. However, both observations and theory indicate that transport from the plasma sheet frequently occurs in the form of localized, transient injections associated with bursty bulk flows (BBFs), also known theoretically as plasma sheet "bubbles" characterized by reduced flux tube entropy ($PV^{5/3}$) where $V$ is the flux tube volume ($\int ds/B$) integrated over the field line, and $P$ is the pressure (Angelopoulos et al., 1992; Pontius & Wolf, 1990).

Using the equilibrium version of the Rice Convection Model (RCM-E), Yang et al. (2015) conducted a systematic investigation of idealized storm main phases. They demonstrated that bubble injections can dominate ring current buildup during intense storms. Their results showed that the bubble contribution increases from ~20% for weak storms to ~50% for moderate storms and saturates near ~61% for intense storms. In contrast, trapped particle contributions decrease correspondingly, and non-bubble transport remains near ~20% regardless of storm strength. These findings provided the first quantitative framework linking BBFs to storm-time ring current energy enhancement.

Observational studies, however, suggest a more moderate role for injections, though this assessment is subject to inherent limitations. Using Van Allen Probes data from the 17 March 2013 storm, Gkioulidou et al. (2014) estimated that small-scale ion injections penetrating inside geosynchronous orbit accounted for roughly ~30% of the total ring current energy gain. However, the authors acknowledged that this figure may represent an underestimate, resulting from a single-spacecraft interpretation, and that extrapolating the occurrence rates of localized events across the entire nightside magnetosphere from a limited orbital track is necessary. More recently, global coupled simulations using the Multiscale Atmosphere-Geospace Environment (MAGE) model found that bubbles contribute at least ~50% of the inward energy flux during strong storms, but that a substantial fraction of this energy is transported back outward by inertial return flows, reducing the net contribution to the ring current region. Specifically, Sciola et al. (2023) utilized the MAGE model to quantify the enthalpy flux traversing a surface at $6R_e$ across the nightside magnetosphere. They observed that while high-speed, entropy-depleted bubbles were the primary drivers of inward transport, the inertial braking of these flows against the rigid inner magnetosphere generated vortices and tailward flows on the flanks of the injection channel. This quantitative analysis revealed that for the total plasma energy transported inward by bubbles, these return flows transported approximately 40% of that amount back outward, significantly dampening the efficiency of energy retention in the ring current.

A key limitation of equilibrium-based models such as RCM-E is the neglect of inertial effects in the magnetohydrodynamic momentum balance. This slow-flow assumption suppresses dynamic processes such as flow braking, overshoot, and oscillatory behavior that are frequently observed in BBFs (Panov et al., 2010, 2013). To address this limitation, Yang et al. (2019) developed the inertialized Rice Convection Model (RCM-I), which approximately incorporates plasma inertia by modifying the field-aligned current closure. RCM-I supports braking oscillations and buoyancy waves following bubble injections and yields more realistic flow velocities than equilibrium formulations (Sadeghzadeh et al., 2021; Yang et al., 2014).

In this study, we quantify the contribution of plasma sheet bubbles to ring current buildup using a test particle backtracking technique applied to three idealized storm simulations conducted with RCM-I. By systematically varying solar wind and ionospheric driving conditions to represent baseline, moderate, and strong storms, we examine how inertial dynamics alter the relative importance of bubble-driven transport compared to trapped and non-bubble plasma sheet populations.

## METHODOLOGY AND SETUP

### RCM-I Simulations

The RCM-I simulation uses a grid resolution of $I_{max} = 200, J_{max} = 144$ in latitude and longitude directions, respectively. To generate the initial distribution for the backtracking simulation, we process the distribution function output exported from the RCM at the designated start time. This is the time to start the backtracking. A Python interface reads the RCM grid geometry and plasma parameters—specifically the plasma flux tube content ($\eta$) and plasma pressure—to reconstruct the phase space density for a given species. The algorithm scans the RCM spatial grid to identify all valid flux tubes located within the ring current region, defined in this experiment as the radial distance between 2.5 and $6.6 R_e$. Given the RCM's grid resolution, this spatial filtering identifies ~ 6,000-7,000 distinct grid points within the region of interest. By weighing the particle population according to $\eta$ and pressure values at these points, the code generates a full initial synthetic population of $N$~ 500,000 macro-particles, consisting of $N_e$~ 100,000 electrons and $N_p$~ 400,000 protons. This specific combination is not arbitrary but is an algorithmic consequence of our pressure-weighted initialization scheme. In the RCM framework, the total plasma pressure is predominantly carried by protons due to the physically higher proton-to-electron temperature ratio ($T_p/T_e \approx 4-6$). Because our initialization algorithm dynamically allocates macro-particles to match the underlying macroscopic pressure of each species, roughly 80% of the total macro-particles are naturally assigned to protons. Furthermore, this 4:1 ratio is computationally optimal for our phase-space grid, as the model allocates 60 energy invariant ($\lambda$) channels to protons compared to only 30 to electrons. The higher number of proton macro-particles ensures that these broader proton energy channels are adequately populated, minimizing statistical noise during the phase-space reconstruction.

We performed three idealized storm simulations representing baseline, moderate, and strong geomagnetic activity. The runs were distinguished by their imposed solar wind and ionospheric driving conditions, summarized in Table 1. The baseline run employed a polar cap potential (PCP) of 50kV, solar wind density ($N_{sw}$) of $5 cm^{-3}$, solar wind velocity ($V_{sw}$) of 400km/s, and IMF $B_z$ of -5nT, producing a minimum Dst of approximately $-30$nT. The moderate run used PCP=150kV, $N_{sw}$=15$cm^{-3}$, $V_{sw}$=500km/s, and IMF $B_z$ of -10nT, yielding Dst $\approx -100$nT. The strong run employed PCP=220kV, $N_{sw}$=50$cm^{-3}$, $V_{sw}$=600km/s, and IMF $B_z$ of -18nT, producing Dst values reaching approximately −180nT. The parameters were chosen to systematically bracket baseline, moderate, and extreme storm conditions. The PCP values are physically consistent with the corresponding solar wind inputs (effectively represented by Boyle's empirical relationship (Boyle et al., 1997)), while the bubble depletion levels are based on previous RCM calibrations. Varying these numbers from small (baseline) to large (strong) profoundly impacts our results, demonstrating how enhanced driving allows bubbles to overcome inertial braking and transition the ring current from a "trapped-dominated" state (92% trapped) to a "bubble-enhanced" state (~43% bubble contribution).

Table 1. Setup parameters for three idealized storm simulations.

| run | storm intensity | Dst $(nT)$ | $N_{sw}$ $(cm^{-3})$ (max) | $V_{sw}$ $(km/s)$ (max) | IMF $B_z$ $(nT)$ (max) | PCP $(kV)$ (max) | time $(h)$ |
|---|---|---|---|---|---|---|---|
| 1 | baseline | $> -30$ | 5 | 400 | $-5$ | 50 | 8 |
| 2 | moderate | $-50$ to $-100$ | 15 | 500 | $-10$ | 150 | 8 |
| 3 | strong | $-150$ to $-200$ | 50 | 600 | $-18$ | 220 | 8 |

Each simulation consisted of three phases. The first phase (≈2 h) resembled a growth phase with stretched magnetic field lines under baseline solar wind conditions (PCP=50kV, $N_{sw}$=5$cm^{-3}$, $V_{sw}$=400km/s, IMF $B_z$ of -5nT). In the second phase (T > 02:00), bubble injections were imposed along the nightside boundary by introducing localized reductions in flux tube entropy. The azimuthal width of each bubble was ~0.2 (in radians), and the degree of depletion gradually increased from 95% toward 50% of the background value. Approximately 30, 45, and 65 bubble injections were imposed in the baseline, moderate, and strong runs, respectively. The final phase (after T > 06:00) restored the solar wind and ionospheric conditions to their initial values, allowing the system to relax.

The RCM-I framework represents a self-consistent coupling between the multi-fluid physics of the Rice Convection Model (RCM) and a three-dimensional magnetohydrodynamic (MHD) solver (Lemon et al., 2003; Silin et al., 2013). Unlike the standard equilibrium configurations, the inertial mode solves the full time-dependent momentum equation, explicitly retaining the inertial terms

$$\rho \frac{d\mathbf{v}}{dt} = \mathbf{J} \times \mathbf{B} - \nabla P \tag{1}$$

where $\rho$ is the plasma density, **v** is the plasma velocity, **J** is the current density, **B** is the magnetic field, and $P$ is the isotropic thermal pressure. In this regime, the artificial frictional damping used in equilibrium version (Toffoletto et al., 2003) is deactivated, allowing the solver to evolve the magnetic field topology dynamically via Faraday's Law ($\partial \mathbf{B}/\partial t = -\nabla \times \mathbf{E}$). The momentum equation is solved across the full 3D magnetospheric space, rather than just the equatorial plane. Additionally, the thermal pressure used in this equation inherently incorporates gradient and curvature drifts because the RCM explicitly calculates and applies these energy-dependent drifts during its plasma advection step. This approach allows the simulation to capture transient magnetospheric phenomena where plasma acceleration and flow braking are dominant, rather than assuming a series of quasi-static equilibria. The coupling logic operates on a synchronized exchange cycle. During each cycle, the RCM calculates the adiabatic drift of distinct plasma species on closed field lines to determine the global distribution of plasma pressure and density. These distributions are mapped onto the MHD grid to serve as source terms for the momentum equation. The MHD solver then advances the plasma bulk velocity and magnetic field configuration over the physical time interval. The updated magnetic field topology—specifically the specific flux tube volumes ($V = \int ds/B$) and field strengths—is fed back to the RCM. This feedback loop modifies the gradient/curvature drifts and the ionospheric conductance, which in turn regulates the self-consistent calculation of the ionospheric electrostatic potential via the conservation of current

$$\nabla . (\mathbf{\Sigma} . \nabla \Phi) = -J_\parallel \sin I \tag{2}$$

where $\Phi$ is the scalar electrostatic potential on the ionospheric grid, $\mathbf{\Sigma}$ represents the conductance tensor, $J_\parallel$ is the current flowing into/out of the ionosphere from the magnetosphere, and $I$ is the magnetic dip angle. By incorporating inertial effects, the RCM-I offers a significant advancement in modeling the dynamic Magnetosphere-Ionosphere (M-I) coupling during geomagnetically active periods. While equilibrium models damp out rapid flows, the inertial mode supports the simulation of high-speed flows, such as BBFs and plasma bubbles. The inertial current per unit flux in the near-equatorial plane is given by

$$j_\parallel^{in} = \zeta \rho V \boldsymbol{e_B} . \nabla B_{eq} \times \frac{d\mathbf{v}_{eq}}{dt} \tag{3}$$

where $\zeta$ is a correction factor for mass distribution, $\rho V$ is mass per unit flux, $\boldsymbol{e_B}$ is the unit vector along magnetic field direction, $\nabla B_{eq}$ is the vector gradient of the equatorial magnetic field strength, and $d\mathbf{v}_{eq}/dt$ is the vector representing plasma acceleration. This allows depleted flux tubes to accelerate earthward and potentially overshoot their equilibrium positions, penetrating deeper into the inner magnetosphere than static pressure balance would predict. This capability is essential for accurately simulating the injection of ring current particles, the generation of inductive electric fields during substorm dipolarization, and the transient response of the ionospheric electrodynamics to magnetospheric driving (Ohtani et al., 2007). Two loss mechanisms are explicitly included in our simulations, and they impact the system differently: (a) electron precipitation: this is actively modeled using the strong pitch-angle scattering limit. It matters significantly in our simulations because auroral electron precipitation directly enhances ionospheric conductance. This updated conductance is crucial as it regulates the self-consistent calculation of the ionospheric electrostatic potential, thereby directly influencing the convection electric fields that drive bubble transport. (b) ion charge exchange: this is actively calculated using an energy-dependent loss rate function for protons. While charge exchange fundamentally matters for ring current decay, its macroscopic impact in these specific idealized runs is negligible. Because our modeled plasma consists solely of protons (omitting faster-decaying oxygen ions), their long charge-exchange lifetimes are too slow to cause noticeable pressure depletion or Dst recovery during the brief 2-hour relaxation window of our simulations. To clarify the relationship between an "RCM bubble injection" and the resulting "MHD flow injection" within the RCM-I framework, it is worth detailing the propagation chain of these low-entropy flow channels. The RCM imposes a local entropy and pressure depletion at its high-latitude (tailward) boundary. This depletion is then written directly into the MHD thermodynamic state. When the MHD solver reads this localized pressure deficit inside the bubble channel, the outward pressure gradient force drops, creating a severe imbalance with the inward magnetic tension force ($\boldsymbol{J} \times \boldsymbol{B}$). Because no kinematic velocity vectors are ever artificially imposed, this unbalanced force dynamically accelerates the plasma, generating a self-consistent, interchange-driven earthward bulk flow entirely from first-principle momentum equations.

**Particle Backtracking**

The particle backtracking simulation reconstructs the historical trajectories of magnetospheric plasma species (electrons and protons) by integrating their bounce-averaged drift velocities backward in time through a sequence of electromagnetic field configurations. The transport logic is grounded in the RCM formalism, where particle motion across the 2D ionospheric grid is driven by gradients in the effective potential ($V_{eff}$), which combines the convection electric potential with an energy-dependent magnetic drift term. The drift equation is

$$\mathbf{v} = \frac{\mathbf{B} \times \nabla V_{eff}}{B^2} \tag{4}$$

where

$$V_{eff}(I, J, t) = \Phi(I, J, t) + s\, \lambda_s\, V(I, J, t)^{-2/3} \tag{5}$$

This potential is calculated at each time step as a combination of the convection electric potential ($\Phi$) and the flux tube volume ($V$) scaled by the particle's energy invariant ($\lambda_s$) and charge sign ($s = +1$ for protons and $s = -1$ for electrons). The electrostatic potential ($\Phi$) incorporates inertial effects because it is derived from a current conservation equation that explicitly includes an inertial source term. Note that $I$ and $J$, respectively, are grid indices along latitude and longitude. To accurately translate these ionospheric drift solutions into physical trajectories in the magnetosphere, the backtracker must account for the dynamically evolving magnetic field topology. If a particle's mapped equatorial position is denoted by $x_e(I, J, t)$, its true velocity in the equatorial plane, $v_{eq}$, is governed by the chain rule

$$v_{eq} = \frac{dx_e}{dt} = \frac{\partial x_e}{\partial I}\frac{dI}{dt} + \frac{\partial x_e}{\partial J}\frac{dJ}{dt} + \frac{\partial x_e}{\partial t} \tag{6}$$

The first two terms represent the physical drift of the plasma across magnetic field lines, driven strictly by the spatial gradients of the effective potential ($\nabla V_{eff}$) computed on the fixed ionospheric grid. The final term, $\partial x_e / \partial t$, represents the purely inductive drift caused by the time-varying magnetic field—defined precisely as the equatorial velocity of a fixed ionospheric grid point. Because our backtracker continually re-evaluates the geometric mapping of the static $(I, J)$ ionospheric footprints to their moving $(X, Y)$ equatorial positions frame-by-frame, this dynamic interpolation inherently and automatically captures the $\partial x_e / \partial t$ inductive motion. Thus, the algorithm successfully resolves the localized stretching, inertial flow braking, and interchange oscillations of the magnetic field lines entirely independent of the $\mathbf{E} \times \mathbf{B}$ cross-field convection. The temporal evolution of particle positions is solved using a fourth-order Runge-Kutta (RK4) integration scheme with adaptive step-size control based on local characteristic length scales to ensure numerical accuracy within the curvilinear coordinate system.

To run the simulation, the tool requires time-dependent field outputs from the RCM. Specifically, it utilizes the time histories of the electric potentials ($\Phi$), flux tube volume ($V$), and geometric grid mappings (colatitude and local time). The tool initializes a distribution of particles defined by their species, initial positions, and adiabatic invariants. Throughout the backtracking process, the logic strictly enforces the RCM's geometric limits: an elliptical boundary mask is applied to the grid, and particles that trace back to positions outside this boundary (the model's separatrix) are flagged as "stopped" or inactive to prevent invalid extrapolation. The primary output of the tool is a comprehensive dataset containing the full phase-space trajectory ($X$, $Y$ coordinates mapped to the equatorial plane), time stamps, and status flags for every particle. Based on their backtracked origin and transport history, particles were categorized as originating from (1) bubble-associated flux tubes, (2) non-bubble plasma sheet flux tubes, or (3) trapped inner magnetospheric populations. Contributions were then quantified in terms of their relative share of ring current pressure inside ($R < 6.6R_e$).

The particle tracing methodology utilized by Yang et al. (2015) was computationally expensive, relying on a deterministic grid-based approach that initialized one test particle for every energy invariant channel at every spatial grid point inside geosynchronous orbit. This resulted in a massive computational load of ~ 8 to 12 million particles per simulation. The primary limitation of this "brute force" method is its inefficiency; it expends equal computational resources tracing regions of phase space with negligible density as it does for the dominant populations, effectively tracing "empty" channels that contribute little to the final ring current pressure. In contrast, our tool employs a *stratified* sampling methodology that maintains physical accuracy while reducing the computational burden by two orders of magnitude. Instead of tracing every fixed channel, we generate a representative population weighted by local entropy and pressure, then downsampled to ~100,000 particles using quantile binning to preserve the shape of the energy invariant ($\lambda$) distribution and species (electrons and protons) ratios. Validated by Kolmogorov-Smirnov (KS) tests, which confirm the statistical similarity between the downsampled subset and the full distribution, this approach captures the macroscopic energy dynamics with accuracy comparable to the larger datasets but with vastly superior computational efficiency and memory management.

It is important to note a specific limitation regarding the treatment of particle losses during the backtracking phase. While the forward RCM-I simulation explicitly includes time-dependent loss mechanisms, these non-adiabatic processes are not inverted during the backward test-particle tracing. The backtracker relies strictly on collisionless, bounce-averaged adiabatic drift equations and does not attempt to mathematically restore particles that were lost prior to the end of the simulation. Therefore, the fractional pressure contributions calculated in this study strictly represent the origins of the *surviving* ring current plasma. As previously noted, macroscopic pressure depletion due to proton charge exchange is negligible over our brief 2-hour simulation window, ensuring this methodological constraint does not significantly impact the final calculated total energy fractions (which are heavily proton-dominated). However, because electrons precipitate on much faster timescales, the backtracked electron distributions inherently omit the previously precipitated populations. Incorporating time-reversed particle loss algorithms into the backtracker to fully account for these missing populations remains an objective for our future work.

## RESULTS

The magnitude of the storm-time Dst index is determined by the synergy between global solar wind drivers, which establish the energy reservoir and convection strength, and localized bubble injections, which facilitate the rapid transport of this energy into the inner magnetosphere. Our simulations using the RCM-I confirm a strong dependence of bubble contribution on geomagnetic storm intensity, while highlighting the limiting role of inertia in ring current buildup. Figure 1 shows estimated Dst for three simulations using the Dessler-Parker-Sckopke (DPS) relation by calculating the total energy within geosynchronous orbit. The use of the DPS relation is highly justified because it aligns with established practices in recent ring current literature, enabling direct, "apples-to-apples" quantitative comparisons between our RCM-I results and prior non-inertial RCM-E benchmarks. Furthermore, since our study focuses strictly on the energy composition within geosynchronous orbit, the DPS formula provides a reliable, linear proxy to translate local plasma energy buildup into estimated Dst values, which perfectly serves our objective of categorizing the idealized simulations into baseline, moderate, and strong storm intensities. A "strong" storm (with Dst < -150nT) results from the combination of a massive plasma source (high $N_{sw}$) and highly depleted bubbles that act as fast conduits to fill the inner magnetosphere, whereas "moderate" or "baseline" cases result from lower densities or "heavier" bubbles that stall further out in the tail, failing to build sufficient ring current pressure to significantly depress the magnetic field. We believe that the absence of Dst recovery during the simulation's relaxation phase is fundamentally driven by the omission of heavy oxygen ($O^+$) ions, which normally facilitate the rapid initial decay of real storms through fast charge-exchange loss. Because the modeled ring current is composed solely of protons, their much longer charge-exchange lifetimes are insufficient to produce noticeable pressure decay within the short two-hour relaxation window. Additionally, when the enhanced storm-time convection is turned off, the massive resident "trapped" population is not swept out of the system but remains in stable closed orbits, persistently maintaining the high total plasma pressure and depressed Dst.

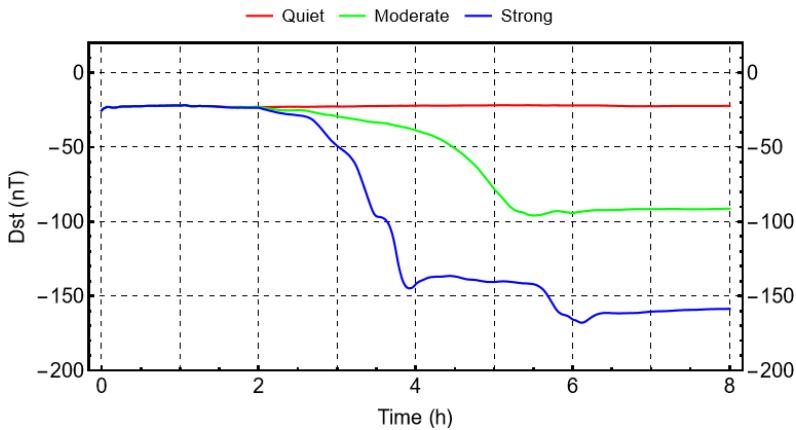

Figure 1. The temporal evolution of Dst index for 3 idealized runs.

We conducted our primary backtracking analysis using a sample size of ~100,000 test particles to optimize computational resources. To validate that this sample size was sufficient to capture the macroscopic physics of the ring current, we performed a sensitivity analysis by repeating another simulation with a larger sample of one million particles (not presented here). A comparison of the results revealed that the 100,000-particle and 1-million-particle stratified samples produced almost identical relative contributions of bubble, non-bubble, and trapped particles to the total pressure inside geosynchronous orbit. This convergence confirms that the 100,000-particle sample provides a statistically robust representation of the system's global energy dynamics at a computational cost lower than that of the full distribution.

Figure 2 shows example snapshots of the entropy and pressure at T=02:35 for run 3 where the geomagnetic activity is strong. Panel (a) shows three clusters of colored dots overlapping on the entropy profile: (i) trapped particles (green) that never hit the RCM boundary during the simulation window, representing the stable, resident ring current population, (ii) bubble particles (red) that arrived at the RCM boundary in coincidence with bubble injection events (shown by black arrows), and (iii) non-bubble particles (blue) that arrived at the RCM boundary outside of discrete injection times/locations. These non-bubble particles represent the slow transport population that fills the region between injection events. They are not drifting from the bubble itself. They represent the background plasma sheet convection. The earthward transport of plasma along with sustained injections of depleted flux tubes resulted in a gradual buildup of particle pressure on the nightside.

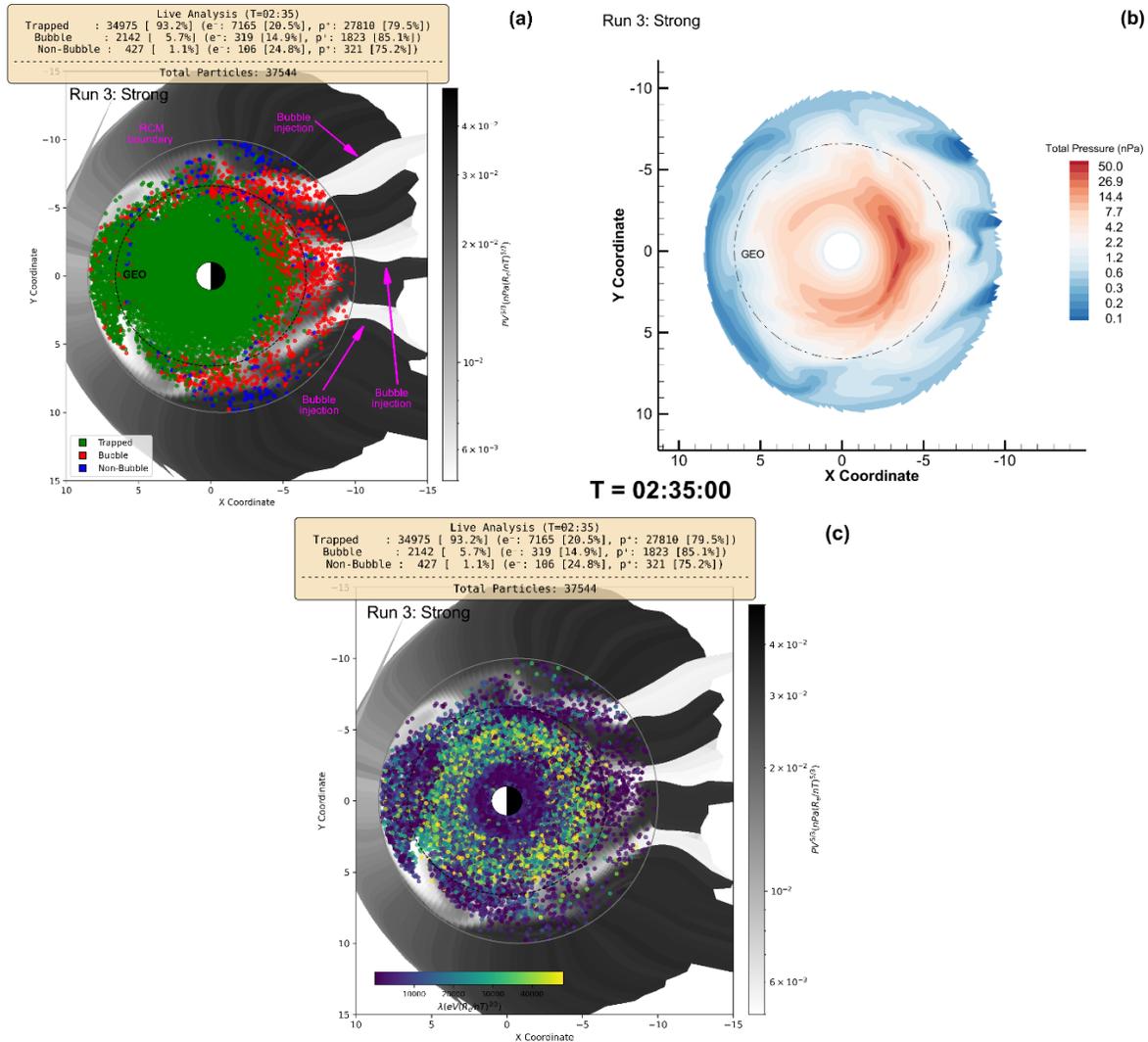

Figure 2. Equatorial plane snapshots of (a) and (c) $PV^{5/3}$ (in units of nPa $(R_E/nT)^{5/3}$) and (b) total plasma pressure (in nPa) at T=02:35 in run 3. In panel (a), test particles are overplotted as green, red, and blue dots denoting trapped, bubble, and non-bubble populations, respectively. The same particles are also color-coded using a Virdis colormap according to their energy invariant $\lambda$ (in $eV(R_e/nT)^{2/3}$) in panel (c). Black arrows indicate injection locations along the boundary. The X and Y coordinates are given in units of Earth radii ($R_e$).

Panel (c) displays the spatial distribution of the same test particles, color-coded by their $\lambda$ values. The particle populations are represented using approximately 90 invariant energy channels spanning from a few to about 50,000 $eV(R_e/nT)^{2/3}$, with 30 channels allocated to electrons and 60 to protons. This panel is essential for understanding the composition of the ring current, whereas panel (a) only explains the source of the volume. Three key features stand out:

First, the innermost region (roughly $R \lesssim 3R_e$) is dominated by particles with a lower energy invariant $\lambda$ (typically $\lambda \lesssim 10{,}000\ eV(R_e/nT)^{2/3}$). This indicates that the deep inner magnetosphere is populated mainly by lower-entropy plasma, consistent with the strong magnetic field compression in this region: because the flux tube volume $V$ is small, even particles with significant kinetic energy correspond to small invariants ($\lambda \approx EV^{2/3}$), while the access of very high-$\lambda$ particles to this depth is restricted by magnetic drift topology. Second, the region between $2R_e < R < 6.6R_e$ contains a broader spread of $\lambda$, including a distinct population with $\lambda \gtrsim 20{,}000\ eV(R_e/nT)^{2/3}$. These high-$\lambda$ (high-energy invariant) particles are primarily associated with the non-bubble convection and portions of the pre-existing trapped plasma sheet, rather than freshly injected bubbles. Third, this center and outermost region ($2 - 6.6R_e$) also contains a prominent low-$\lambda$ population that pertains directly to the bubbles. This is because the bubble channels themselves introduce these particles with comparatively lower $\lambda$ (often below $\sim 10{,}000\ eV(R_e/nT)^{2/3}$), reflecting the defining property of bubbles as entropy-depleted flux tubes ($PV^{5/3}$). It is precisely this low-entropy state that makes them interchange-unstable, allowing them to penetrate deep into the dipole field where high-entropy flux tubes cannot reach. Thus, panel (c) supports the interpretation that bubble injections are the primary mechanism for the inward delivery of low-$\lambda$ plasma which subsequently generates the dominant storm-time pressure through deep adiabatic compression, whereas the high-$\lambda$ populations reflect the slower, non-bubble convection pathways. For completeness and to allow readers to observe the intensity-dependent scaling of these transport mechanisms, the corresponding entropy and phase-space snapshots for the baseline and moderate cases are provided in the Appendix.

To quantify the contributions of different transport mechanisms to the ring current pressure, we employed a reverse-trajectory analysis using the electromagnetic fields generated by the RCM-I simulation. As mentioned, the test particles were initialized at the time of interest (e.g., the end of the simulation) at grid points distributed throughout the ring current region ($R < 6.6R_e$). To ensure the particle ensemble accurately represented the macroscopic fluid state, each test particle was assigned a partial pressure weight derived directly from the RCM-calculated plasma pressure at its initialization grid point. These particles were then traced backward in time to determine their origin, specifically whether they entered the system from the RCM boundary or resided within the inner magnetosphere throughout the event. The classification of particles as trapped, bubble, or non-bubble was determined by their boundary exit behavior during the backtracking process. Finally, the relative contribution of each transport mode was calculated by aggregating the partial pressure weights of the particles in each category. The total pressure contribution for a specific fate (e.g., bubble) was computed as the sum of the pressures of all particles assigned to that fate, which was then normalized by the total pressure of all particles in the region of interest to yield a percentage. This partial pressure weighting, combined with the Lagrangian tracking scheme, allows for a precise decomposition of the macroscopic ring current energy density, linking the final pressure accumulation directly to the specific transport history of the plasma parcels.

Using the particle backtrack simulation, we found that in the baseline run (with Dst $\approx -30$nT), the ring current was dominated by trapped particles (~92%), with bubble-related transport contributing only ~5% and non-bubble plasma sheet sources accounting for the remaining few percent. As storm intensity increased, bubble contributions rose substantially: to ~35% for the moderate storm (Dst $\approx -100$nT) and to ~43% for the strong storm (Dst $\approx -180$nT). Correspondingly, the trapped population decreased from ~92% in the baseline case to ~43% in the strong storm, while the non-bubble plasma sheet contribution increased modestly to ~14–15%. These trends are qualitatively consistent with equilibrium RCM-E results, which predict increasing dominance of bubble-driven transport with storm strength. While the absolute bubble contribution in our RCM-I simulations saturates near ~40%–significantly lower than the ~61% reported in RCM-E–this difference is physically reconciled by the "inertial penalty". Figure 3 compares the predicted energy contribution in percentage to the ring current between RCM-E and RCM-I for a storm with a minimum Dst of ~ -150nT. RCM-E assumes depleted flux tubes penetrate and remain without significant rebound. However, as demonstrated by Sciola et al. (2023), the braking of these flows generates return flows that transport roughly 40% of the inflowing energy back outward. Applying this 40% loss factor to the RCM-E potential (61% × 0.6) yields ~36.6%, aligning well with our RCM-I findings.

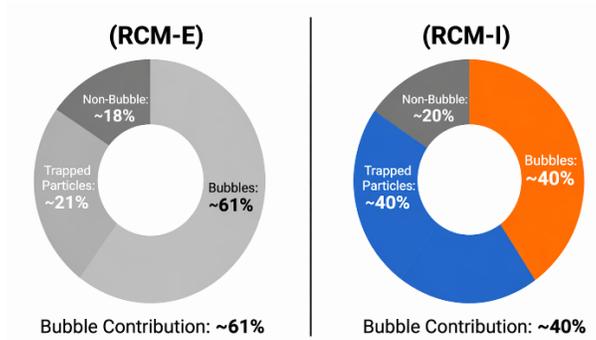

Figure 3. Estimated ring current energy partitioning during strong storms in the RCM-E (left) and RCM-I (right) studies.

Yang et al. (2015) originally hypothesized that including inertial effects would enable plasma sheet bubbles to penetrate deeper into the inner magnetosphere, potentially increasing their contribution beyond the ~61% predicted by equilibrium models. Using RCM-I, Yang et al. (2019) revised this expectation. Their results showed that inertia does not simply enhance earthward transport; instead, it introduces braking (interchange) oscillations and buoyancy waves that radiate energy away from the injection channel. These inertial dynamics significantly slow bubble propagation relative to non-inertial models and redistribute part of the injected energy into vortical flows and wave activity, effectively acting as a brake on ring current buildup rather than an accelerator.

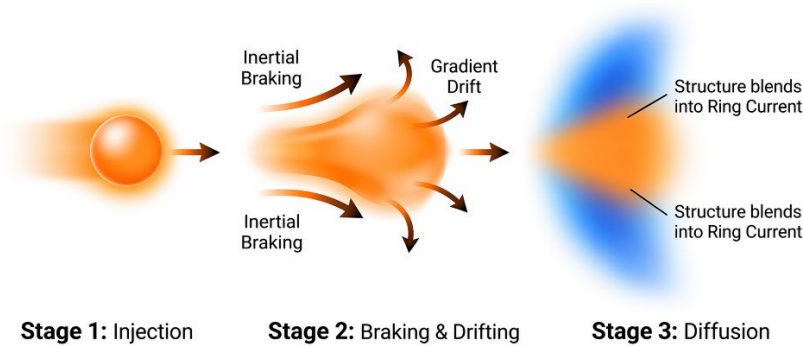

Figure 4. Schematic illustration of the evolution of bubbles in RCM-I, showing the tendency of gradient and curvature drift to diffuse the bubbles into the background.

As bubbles approach the stronger inner magnetospheric field, inertial braking allows gradient and curvature drifts to act over longer timescales, diffusing the depleted flux tubes before they can penetrate deeply or remain coherent. This reduces the efficiency with which bubbles displace the pre-existing plasma population, leaving a substantial trapped component even during strong storms. In contrast, equilibrium models such as RCM-E, which neglect inertia, tend to preserve more coherent bubble structures and therefore predict deeper penetration and higher bubble contributions (see Figure 4).

We repeated the strong storm run with the inertial term switched off. This increased the bubble contribution from ~43% (RCM-I) to ~48% (RCM-E), while leaving the trapped population high at ~48%. This directly confirms the hypothesis derived from Sciola et al. (2023) and Yang et al. (2019) that inertia acts as a brake on ring current buildup. A major insight from this experiment is that the bubble contribution didn't restore to the ~61% benchmark and was capped at ~48%. This is not solely due to inertial physics and implies a difference in boundary conditions as Yang et al. (2015) tuned boundary conditions to force the model to match Geotail velocity statistics. There are several points worth noting here.

1. Within the RCM framework, the Dst index is not derived from particle tracking or percentage contributions. Instead, it is calculated directly using the DPS relation. We calculate the absolute total energy in the inner magnetosphere (integrating the flux tube volume content, energy invariant, and flux tube volume over the spatial grid). Then we obtain the Dst strictly as a function of this absolute total energy (i.e., Dst $\propto$ −total energy). This means that Dst only sees the absolute volume of plasma pressure built up, regardless of whether those particles originated from bubbles, background convection, or pre-existing trapped populations.
2. The difference in how this "total energy" accumulates stems from the fundamentally different partial differential equations solved by the two models. RCM-E assumes a massless plasma. The momentum equation ignores inertia, and field-aligned currents are driven instantly by pressure gradients. Because there is zero "friction," injected bubbles penetrate deeply and smoothly, efficiently pushing the pre-existing ring current plasma out of the way. However, RCM-I solves the full time-dependent momentum equation, explicitly retaining the inertial term ($\rho\, d\mathbf{v}/dt$). This introduces a large inertial field-aligned current into the ionospheric potential solver.
3. When identical initial and boundary conditions are applied, these equations dictate entirely different magnetic and kinematic responses. In RCM-I, the inertial term mathematically acts as a brake on the incoming flows. The large background plasma resists displacement, meaning the resident "trapped" population is not swept out, but instead remains in stable closed orbits. The incoming high-speed flows violently "pile up" against this trapped population and the stronger inner magnetic field, driving a significant adiabatic pressure enhancement. This large absolute pressure buildup translates into substantial "total energy," driving Dst down to ~-200nT. In contrast, in RCM-E, the lack of inertial braking allows the smooth displacement and loss of the resident plasma out of the dayside magnetopause. Without the intense pile-up, the absolute "total energy" integrated over the grid remains modest, resulting in a significantly weaker Dst drop.
4. The ~48% vs ~43% statistic is merely a "paradox of proportions" and has no mathematical bearing on the absolute Dst drop, which is purely a measure of total absolute pressure. A detailed, logical breakdown of the differences between RCM-I and RCM-E results are provided in the Appendix.
5. We deliberately chose not to tune our boundary conditions to match Geotail velocity statistics in order to maintain a strictly controlled experiment. Because the Geotail dataset represents an 11-year statistical average over mostly non-storm conditions, forcing an idealized strong storm to conform to this baseline is physically mismatched. More importantly, tuning the degree of bubble depletion to achieve specific velocities in both an inertial and non-inertial framework would require applying different boundary inputs for each run. By keeping the boundary driving identical, we successfully isolated the physical effect of inertial braking without convoluting it with variable input energies, thereby speculating that the "washout" of the old ring current (predicted by Yang et al., 2015) may have been an artifact of the specific equilibrium tuning, and that in a more realistic inertial regime, the ring current is a hybrid mix of comparable parts, i.e., fresh bubble injection (~43%) and compressed resident plasma (~43%).

## DISCUSSION AND CONCLUSION

We have quantified the contribution of plasma sheet bubbles to storm-time ring current buildup using test particle backtracking applied to three idealized RCM-I simulations representing baseline, moderate, and strong geomagnetic activity. Our results confirm that the relative importance of bubbles increases with storm intensity, in agreement with earlier equilibrium-based modeling. However, the inclusion of inertial effects fundamentally limits the net contribution of bubbles to approximately 40% of the total ring current energy inside ($R < 6.6R_e$), even for intense storms. To accurately compare these results with flux-based studies, we must isolate the newly transported populations from the pre-existing "trapped" reservoir. In our strong storm run, if we consider only the new plasma (bubbles ~40% and non-bubbles ~15%), the relative contribution of bubbles to the inward transport flux becomes 40/(40+15)~73%. This standardized metric falls squarely within the 65-85% range reported by Cramer et al. (2017) and exceeds the 50% lower bound reported by Sciola et al. (2023). Our RCM-I results provide a new lens through which to view observational estimates. Gkioulidou et al. (2014) attributed ~30% of total energy gain to bubbles, leaving ~70% to quasi-steady convection or missed events. However, our finding that bubbles account for ~73% of the actual transport suggests the 30% observational figure is a significant underestimate. This discrepancy likely arises from the single-spacecraft limitations acknowledged by Gkioulidou et al., which would naturally miss spatially localized, transient injections. By modeling the global system, RCM-I suggests that the real contribution of discrete injections to energy gain is likely double the current observational lower bounds.

In conclusion, this analysis synthesizes the existing literature into a coherent physical narrative. We find that equilibrium models (RCM-E) overestimate final energy accumulation (~61%) because they neglect the inertial "return flow penalty". Conversely, global flux models correctly identify bubbles as the dominant transport mechanism (~65–85%) but often overlook the key role of the resident "trapped" energy density. Our RCM-I study bridges these gaps: it captures the high transport dominance of bubbles (~73% of new plasma) while simultaneously accounting for the inertial braking that reduces net energy retention to ~40% and the adiabatic compression of the trapped population. Thus, our results do not contradict the previous 50% or 30% benchmarks; rather, they provide insight into how high-transport rates result in moderate final pressure contributions. Modeling a specific event, such as the 17 March 2013 storm studied by Gkioulidou et al. (2014) and Sciola et al. (2023), would allow for a direct quantitative comparison between our simulation results and in situ spacecraft data. This would verify if the "inertial braking" mechanism in RCM-I correctly explains the discrepancy between the high contribution predicted by equilibrium models (~61%) and the lower contribution (~30%) inferred from observations.

# APPENDIX

This section provides additional phase-space visualizations for the idealized baseline (Run 1) and moderate (Run 2) storm simulations discussed in the main text. For completeness and to allow readers to observe the intensity-dependent scaling of these transport mechanisms, the corresponding entropy and phase-space snapshots for the baseline and moderate storm cases are provided here. These snapshots complement Figure 2 of the main manuscript (which illustrates the strong storm scenario, Run 3) and demonstrate how the penetration depth and composition of plasma sheet bubbles scale with the intensity of the solar wind and ionospheric driving. Furthermore, this document presents a detailed comparison between the equilibrium (RCM-E) and inertial (RCM-I) formulations. It explains how the inclusion of explicit plasma inertia fundamentally alters the magnetosphere-ionosphere coupling equations, resulting in significantly different equipotential patterns, pressure buildups, and Dst index drops even when identical boundary injection profiles are used.

In the baseline storm scenario, the system is driven by baseline solar wind conditions ($N_{sw} = 5 cm^{-3}$, $V_{sw} = 400 km/s$, $IMF\ B_z = -5nT$) and a Polar Cap Potential (PCP) of $50kV$, yielding a minimum Dst of approximately $-30nT$. Under this weak driving, the background convection is slow, and the injected bubbles tend to stall further out in the magnetotail rather than penetrating deep into the inner magnetosphere. As a result, the ring current region remains overwhelmingly dominated by the pre-existing trapped population (~92%), with bubbles contributing only ~5% to the total pressure (Figure A1) .

In the moderate storm scenario, the driving conditions are enhanced ($N_{sw} = 15 cm^{-3}$, $V_{sw} = 500 km/s$, $IMF\ B_z = -10nT$), producing a minimum Dst of approximately $-100nT$. The increased convection and stronger boundary driving allow the injected low-entropy bubbles to overcome inertial braking more effectively than in the baseline case. This results in a deeper penetration of bubble plasma into the $R < 6.6 R_e$ region, with the bubble contribution to the total ring current pressure rising to ~35% (Figure A2).

The fundamental differences in the equipotential patterns between the equilibrium (RCM-E) and inertial (RCM-I) modes stem directly from how each framework handles plasma mass. In the RCM-E configuration, the plasma is assumed to be massless. Consequently, the model solves the standard elliptic magnetosphere-ionosphere coupling equation ($\nabla . \mathbf{J} = 0$) where the electric field instantly reacts and equilibrates to macroscopic pressure gradients.

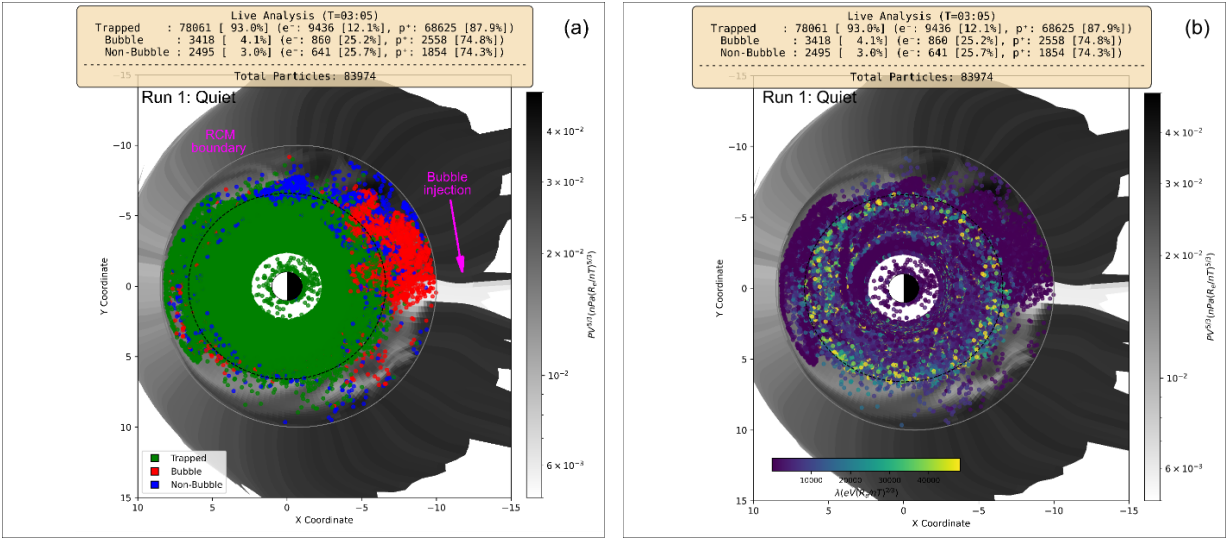

Figure A1. Equatorial plane snapshots for the baseline storm simulation (Run 1). (a) The entropy parameter $PV^{5/3}$ (in units of nPa $(R_E/nT)^{5/3}$) with test particles overplotted as green, red, and blue dots denoting trapped, bubble, and non-bubble populations, respectively. The pink arrow indicates injection location along the boundary. (b) The same spatial distribution of test particles color-coded using a Viridis colormap according to their energy invariant $\lambda$ (in $eV(R_e/nT)^{2/3}$). Compared to the strong storm case in the main text, the injected bubble population (red) fails to penetrate deeply, leaving the low-$\lambda$ inner region almost exclusively populated by the resident trapped (green) plasma.

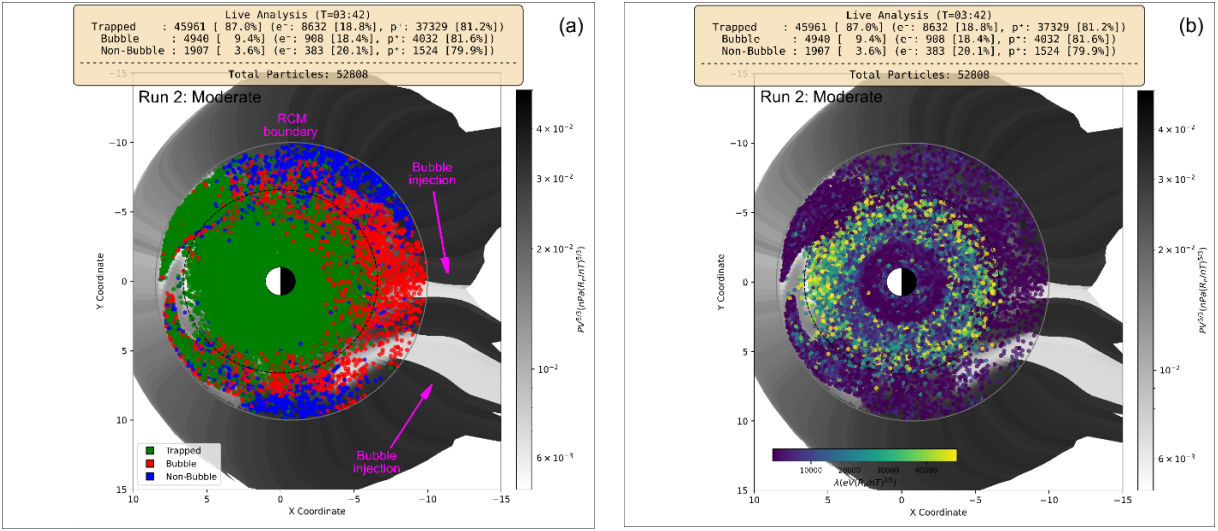

Figure A2. Equatorial plane snapshots for the moderate storm simulation (Run 2). (a) The entropy parameter $PV^{5/3}$ (in units of nPa $(R_E/nT)^{5/3}$) with test particles overplotted as green, red, and blue dots denoting trapped, bubble, and non-bubble populations, respectively. (b) The same test particles color-coded according to their energy invariant $\lambda$. The snapshots illustrate an intermediate state of transport: bubbles penetrate significantly deeper than in the baseline case (Figure A1) and introduce a broader spread of low-$\lambda$ plasma into the inner region, but they do not displace the resident trapped plasma as extensively as they do under strong storm driving.

Since field-aligned currents are driven solely by pressure gradients without inertial lag, the equipotential contours remain smooth, representing an instantaneous, stable force balance throughout the magnetosphere. Conversely, the RCM-I mode introduces explicit plasma inertia into the current continuity equations. The massive plasma resists being instantly accelerated by new drifts when dynamic events, such as random bubble injections, perturb the inner magnetosphere. This inertial resistance generates intense, localized polarization currents that delay the electric field's response and cause the plasma to "slosh." It is this time-dependent inertial delay and the resulting complex vortices that shatter the smooth, large-scale equipotential contours, producing the highly disrupted and noisy patterns characteristic of the RCM-I runs (panels (a) and (c) in Figure A3).

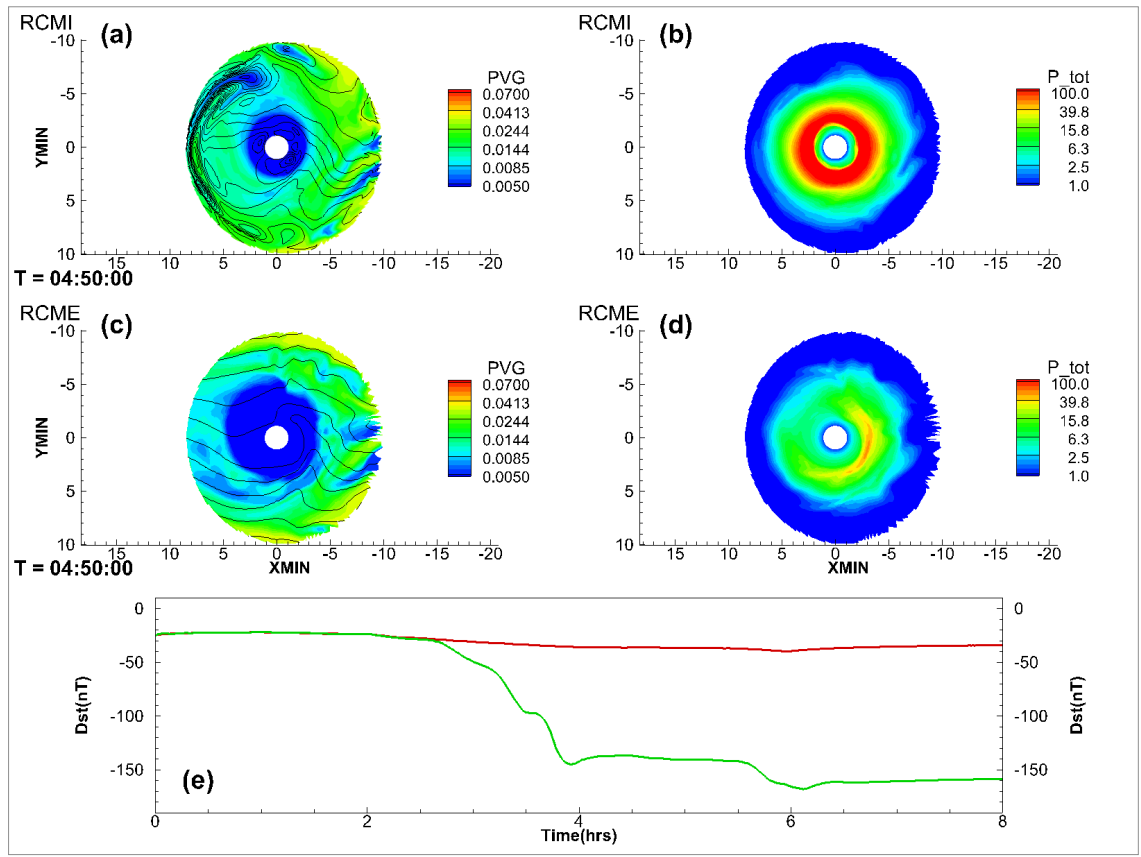

Figure A3. Equatorial plane snapshots comparing the equilibrium (RCM-E, bottom row) and inertial (RCM-I, top row) model configurations at T=04:50:00. The left column displays the plasma flux tube entropy ($PV^{5/3}$) (in nPa($R_E$/nT)$^{5/3}$) overlaid with electrostatic equipotential contours (black lines). The right column shows the corresponding total plasma pressure in nPa. Panel (e) shows the temporal evolution of the estimated Dst index, demonstrating that the massive absolute pressure buildup in the RCM-I drives a significantly deeper Dst drop (~200 nT) than the RCM-E configuration under identical boundary driving.

To ensure that the intense electric fields, plasma sloshing, and inertial return flows observed in the RCM-I simulations are entirely driven by the internal dynamics of the bubble injections, we intentionally applied a steady polar cap potential at the high-latitude boundary. While a more realistic, fluctuating boundary condition could be derived from high-cadence solar wind data to ensure a continuously non-zero boundary derivative ($\partial\Phi/\partial t \neq 0$), doing so would introduce external electrostatic and inductive noise into the system. By maintaining a fixed boundary, we establish a strictly controlled experiment. This guarantees that the highly disrupted internal electric fields we track are unequivocally the result of internal inertial flow braking, completely isolated from external solar wind fluctuations.

In the RCM-E model, injected depleted bubbles (bursty bulk flows) penetrate deep into the inner magnetosphere with almost zero "friction," allowing the surrounding background plasma to smoothly slide out of the way. However, in RCM-I, the background plasma resists this displacement, causing the high-speed earthward flows to violently "pile up" and compress against the stronger magnetic field regions. Because the model calculates the Dst index directly from the total ring current energy (Dst $\propto$ −total energy), this adiabatic pile-up in RCM-I drives a significantly deeper Dst drop (~200nT) than the inertia-free RCM-E run (panel (e) in Figure A3). Consequently, the relatively weak Dst drop observed in the strong RCM-E simulation is a direct result of applying identical initial conditions, boundary conditions, and bubble injection profiles to both models to strictly isolate the effects of inertia, rather than artificially tuning the RCM-E inputs to force a deep Dst drop. It is worth noting that observational evidence shows that prolonged, intense solar wind driving does not always lead to a Dst drop in reality. For example, Tsurutani et al. (2003) analyzed magnetic storms induced by magnetic clouds featuring extreme, prolonged solar wind electric fields, yet found unexpectedly small Dst minimums (e.g., −50 nT and −82 nT). They suggested that a lack of substorm dipolarizations or the rapid entry of low-density plasma prevented sufficient pressure buildup. Similarly, Le et al. (2020) demonstrated that extreme convective driving can produce unexpectedly weak Dst drops if the solar wind dynamic pressure remains low. Thus, the weak Dst produced by the RCM-E under strong driving is not a non-physical artifact, but rather a reflection of how the magnetosphere might respond when the specific mechanisms required for massive pressure buildup—such as inertial pile-up—are absent and the boundary conditions are not explicitly tuned to compensate.

Comparing the ~48% contribution in RCM-E to the ~43% in RCM-I is a paradox of proportions. While bubbles in RCM-E account for a higher percentage of the ring current pressure, they are making up 48% of a vastly smaller absolute total pressure. In RCM-I, the total ring current pressure is massive due to the aforementioned inertial pile-up. The bubbles bring in tremendous absolute pressure, but because the background pressure has also piled up so significantly, the bubbles only account for ~43% of that huge total. Therefore, each bubble in RCM-E may penetrate deeply, but without the inertial pile-up mechanism, its absolute pressure transfer remains comparatively small. Finally, turning the inertia off entirely alters the configuration of the magnetic field topology, shielding efficiency, and pressure buildup. Because RCM-E instantly shields the inner magnetosphere, it fundamentally changes how injected plasma is trapped and transported compared to the delayed, polarization-driven RCM-I. Consequently, applying the exact same boundary conditions and injection profiles to both models does not guarantee a similar Dst response. Directly comparing their absolute pressure buildups or percentage contributions is not useful without acknowledging that these two frameworks govern plasma transport and trapping through entirely different physical laws.

## ACKNOWLEDGMENTS

The work was supported by NASA TMS grant 80NSSC20K1276 and THEMIS contract number NAS5-02099 from UC Berkeley.

# REFERENCES

Angelopoulos, V., Baumjohann, W., Kennel, C. F., Coroniti, F. V., Kivelson, M. G., Pellat, R., Walker, R. J., Lühr, H., & Paschmann, G. (1992). Bursty bulk flows in the inner central plasma sheet. Journal of Geophysical Research: Space Physics, 97(A4), 4027–4039. https://doi.org/10.1029/91JA02701

Boyle, C. B., Reiff, P. H., & Hairston, M. R. (1997). Empirical polar cap potentials. Journal of Geophysical Research: Space Physics, 102(A1), 111–125. https://doi.org/10.1029/96JA01742

Cramer, W. D., Raeder, J., Toffoletto, F. R., Gilson, M., & Hu, B. (2017). Plasma sheet injections into the inner magnetosphere: Two-way coupled OpenGGCM-RCM model results. Journal of Geophysical Research: Space Physics, 122(5), 5077–5091. https://doi.org/10.1002/2017JA024104

Gkioulidou, M., Ukhorskiy, A. Y., Mitchell, D. G., Sotirelis, T., Mauk, B. H., & Lanzerotti, L. J. (2014). The role of small-scale ion injections in the buildup of Earth's ring current pressure: Van Allen Probes observations of the 17 March 2013 storm. Journal of Geophysical Research: Space Physics, 119(9), 7327–7342. https://doi.org/10.1002/2014JA020096

Le, G.-M., Liu, G.-A., & Zhao, M.-X. (2020). Dependence of Major Geomagnetic Storm Intensity (Dst< -100nT) on Associated Solar Wind Parameters. *Solar Physics*, *295*(8), 108. https://doi.org/10.1007/s11207-020-01675-3

Lemon, C., Toffoletto, F., Hesse, M., & Birn, J. (2003). Computing magnetospheric force equilibria. Journal of Geophysical Research: Space Physics, 108(A6), 2002JA009702. https://doi.org/10.1029/2002JA009702

Ohtani, S., Korth, H., Brandt, P. C., Blomberg, L. G., Singer, H. J., Henderson, M. G., Lucek, E. A., Frey, H. U., Zong, Q., Weygand, J. M., Zheng, Y., & Lui, A. T. Y. (2007). Cluster observations in the inner magnetosphere during the 18 April 2002 sawtooth event: Dipolarization and injection at r = 4.6 R E . Journal of Geophysical Research: Space Physics, 112(A8), 2007JA012357. https://doi.org/10.1029/2007JA012357

Panov, E. V., Nakamura, R., Baumjohann, W., Angelopoulos, V., Petrukovich, A. A., Retinò, A., Volwerk, M., Takada, T., Glassmeier, K. -H., McFadden, J. P., & Larson, D. (2010). Multiple overshoot and rebound of a bursty bulk flow. Geophysical Research Letters, 37(8), 2009GL041971. https://doi.org/10.1029/2009GL041971

Panov, E. V., Kubyshkina, M. V., Nakamura, R., Baumjohann, W., Angelopoulos, V., Sergeev, V. A., & Petrukovich, A. A. (2013). Oscillatory flow braking in the magnetotail: THEMIS statistics. Geophysical Research Letters, 40(11), 2505–2510. https://doi.org/10.1002/grl.50407

Pontius, D. H., & Wolf, R. A. (1990). Transient flux tubes in the terrestrial magnetosphere. Geophysical Research Letters, 17(1), 49–52. https://doi.org/10.1029/GL017i001p00049

Sadeghzadeh, S., Yang, J., Wang, C., Mousavi, A., Wang, W., Sun, W., Toffoletto, F., & Wolf, R. (2021). Effects of Bubble Injections on the Plasma Sheet Configuration. Journal of Geophysical Research: Space Physics, 126(6), e2021JA029127. https://doi.org/10.1029/2021JA029127

Sciola, A., Merkin, V. G., Sorathia, K., Gkioulidou, M., Bao, S., Toffoletto, F., Pham, K., Lin, D., Michael, A., Wiltberger, M., & Ukhorskiy, A. (2023). The Contribution of Plasma Sheet Bubbles to Stormtime Ring Current Buildup and Evolution of Its Energy Composition. Journal of Geophysical Research: Space Physics, 128(11), e2023JA031693. https://doi.org/10.1029/2023JA031693

Silin, I., Toffoletto, F., Wolf, R., & Sazykin, S. (2013). Calculation of Magnetospheric Equilibria and Evolution of Plasma Bubbles with a New Finite-Volume MHD/Magnetofriction Code (Vol. 2013, pp. SM51B-2176). Presented at the AGU Fall Meeting Abstracts. Retrieved from https://ui.adsabs.harvard.edu/abs/2013AGUFMSM51B2176S

Toffoletto, F., Sazykin, S., Spiro, R., & Wolf, R. (2003). Inner magnetospheric modeling with the Rice Convection Model. Space Science Reviews, 107(1/2), 175–196. https://doi.org/10.1023/A:1025532008047

Tsurutani, B. T., Zhou, X.-Y., & Gonzalez, W. D. (2003). A lack of substorm expansion phases during magnetic storms induced by magnetic clouds. In A. Surjalal Sharma, Y. Kamide, & G. S. Lakhina (Eds.), *Geophysical Monograph Series* (Vol. 142, pp. 23–36). Washington, D. C.: American Geophysical Union. https://doi.org/10.1029/142GM03

Yang, J., Wolf, R. A., Toffoletto, F. R., Sazykin, S., & Wang, C. (2014). RCM-E simulation of bimodal transport in the plasma sheet. Geophysical Research Letters, 41(6), 1817–1822. https://doi.org/10.1002/2014GL059400

Yang, J., Toffoletto, F. R., Wolf, R. A., & Sazykin, S. (2015). On the contribution of plasma sheet bubbles to the storm time ring current. Journal of Geophysical Research: Space Physics, 120(9), 7416–7432. https://doi.org/10.1002/2015JA021398

Yang, J., Wolf, R., Toffoletto, F., Sazykin, S., Wang, W., & Cui, J. (2019). The Inertialized Rice Convection Model. Journal of Geophysical Research: Space Physics, 124(12), 10294–10317. https://doi.org/10.1029/2019JA026811